\documentclass{ieeetj}
\usepackage{cite}
\usepackage{amsmath,amssymb,amsfonts}
\usepackage{algorithmic}
\usepackage{graphicx,color}
\usepackage{textcomp}
\usepackage{xcolor}
\usepackage{hyperref}
\hypersetup{hidelinks=true}
\usepackage{algorithm,algorithmic}
\def\BibTeX{{\rm B\kern-.05em{\sc i\kern-.025em b}\kern-.08em
    T\kern-.1667em\lower.7ex\hbox{E}\kern-.125emX}}
\AtBeginDocument{\definecolor{tmlcncolor}{cmyk}{0.93,0.59,0.15,0.02}
\definecolor{NavyBlue}{RGB}{0,86,125}}
\graphicspath{{./}}

\usepackage{accents}
\usepackage{cases}  
\usepackage{algorithm}      
\usepackage{algorithmic}      
\usepackage{eqparbox}
\usepackage{xargs}   
\usepackage{spconfpdkul}

\usepackage{todonotes}
\usepackage{pgfplots}   
\usepackage{ifthen}
\usepackage[acronym, shortcuts]{glossaries}  
\usepackage{makecell}
\usepackage{subcaption}  
\usepackage[activate={true,nocompatibility},final]{microtype}  
\makeatletter
\g@addto@macro\normalsize{%
  \setlength\abovedisplayskip{1pt plus 1pt minus 1pt}%
  \setlength\belowdisplayskip{1pt plus 1pt minus 1pt}%
  \setlength\abovedisplayshortskip{0pt plus 1pt}%
  \setlength\belowdisplayshortskip{0pt plus 1pt minus 1pt}%
}
\makeatother
\newacronym{wasn}{WASN}{wireless acoustic sensor network}
\newacronym{danse}{DANSE}{distributed adaptive node-specific signal estimation}
\newacronym{rsdanse}{rS-DANSE}{relaxed simultaneous node-updating distributed adaptive node-specific signal estimation}
\newacronym{dansep}{TI-DANSE$^+$}{topology-independent distributed adaptive node-specific signal estimation PLUS[TO BE CHANGED]}
\newacronym{idanse}{iDANSE}{iterationless distributed adaptive node-specific signal estimation}
\newacronym{ti}{TI}{topology-independent}
\newacronym{tiidanse}{TI-iDANSE}{topology-independent iterationless distributed adaptive node-specific signal estimation}
\newacronym{idansep}{iDANSE$^+$}{iterationless distributed adaptive node-specific signal estimation PLUS[TO BE CHANGED]}
\newacronym{gevddansep}{GEVD-DANSE$^+$}{GEVD-based distributed adaptive node-specific signal estimation PLUS[TO BE CHANGED]}
\newacronym{gevd_or_dansep}{(GEVD-)DANSE$^+$}{(GEVD-based) distributed adaptive node-specific signal estimation PLUS[TO BE CHANGED]}
\newacronym{tdanse}{T-DANSE}{tree-DANSE}
\newacronym{dasf}{DASF}{distributed adaptive signal fusion}
\newacronym{dansf}{DANSF}{distributed adaptive node-specific signal fusion}
\newacronym{tidanse}{TI-DANSE}{topology-independent distributed adaptive node-specific signal estimation}
\newacronym{tigevddanse}{TI-GEVD-DANSE}{topology-independent GEVD-based distributed adaptive node-specific signal estimation}
\newacronym{ti_or_gevddanse}{TI-(GEVD-)DANSE}{topology-independent (GEVD-based) distributed adaptive node-specific signal estimation}
\newacronym{gevddanse}{GEVD-DANSE}{GEVD-based distributed adaptive node-specific signal estimation}
\newacronym{gevd}{GEVD}{generalized eigenvalue decomposition}
\newacronym{gevc}{GEVC}{generalized eigenvector}
\newacronym{gevl}{GEVL}{generalized eigenvalue}
\newacronym{lmmse}{LMMSE}{linear minimum mean square error}
\newacronym{mmse}{MMSE}{minimum mean square error}
\newacronym{mwf}{MWF}{multichannel Wiener filter}
\newacronym{gevdmwf}{GEVD-MWF}{GEVD-based multichannel Wiener filter}
\newacronym[longplural={spatial covariance matrices}]{scm}{SCM}{spatial covariance matrix}
\newacronym{snr}{SNR}{signal-to-noise ratio}
\newacronym{mse}{MSE}{mean square error}
\newacronym{tad}{TAD}{target activity detector}
\newacronym{vad}{VAD}{voice activity detector}
\newacronym{spp}{SPP}{speech presence probability}
\newacronym{sad}{SAD}{source activity detector}
\newacronym{mst}{MST}{minimum spanning tree}
\newacronym{spt}{SPT}{shortest-path tree}
\newacronym{gsbcd}{GSBCD}{Gauss-Seidel block-coordinate descent}
\newacronym{ao}{AO}{alternating optimization}
\newacronym{mmut}{MMUT}{multiple max-$|\Uk^i|$ trees}
\newacronym{mc}{MC}{Monte-Carlo}
\newacronym{ac}{AC}{algebraic connectivity}
\newacronym{fc}{FC}{fully connected}
\newacronym{stft}{STFT}{short-time Fourier transform}
\newacronym{dft}{DFT}{discrete Fourier transform}
\newacronym{mil}{WMI}{Woodbury matrix identity}
\newacronym{poss}{POSS}{partially overlapping signal subspace}
\newacronym{foss}{FOSS}{fully overlapping signal subspace}
\newacronym{dof}{DoF}{degree of freedom}
\newacronym{dmwf}{dMWF}{distributed multichannel Wiener filter}
\newacronym{gevddmwf}{GEVD-dMWF}{GEVD-based distributed multichannel Wiener filter}
\newacronym{tidmwf}{TI-dMWF}{topology-independent distributed multichannel Wiener filter}
\newacronym{tigevddmwf}{TI-GEVD-dMWF}{topology-independent GEVD-based distributed multichannel Wiener filter}
\newacronym{tidmwfp}{TI-dMWF$^+$}{topology-independent distributed multichannel Wiener filter PLUS[TO BE CHANGED]}
\newacronym{ti_or_dmwf_or_p}{(TI-)dMWF($^+$)}{(topology-independent) distributed multichannel Wiener filter (PLUS[TO BE CHANGED])}
\newacronym{lcmv}{LCMV}{linearly constrained minimum variance}
\newacronym{gsc}{GSC}{generalized sidelobe canceller}
\newacronym{mvdr}{MVDR}{minimum variance distortionless response}
\newacronym{bf}{BF}{beamformer}
\newacronym{rtf}{RTF}{relative transfer function}
\newacronym{wlog}{w.l.o.g.}{without loss of generality}
\newacronym{stoi}{STOI}{short-term objective intelligibility}
\newacronym{ism}{ISM}{image-source method}
\newacronym{ser}{SER}{signal-to-error ratio}
\newacronym{gls}{GLS}{global-local subspaces}
\newacronym{cgls}{cGLS}{constrained global-local subspaces}
\newacronym{fos}{FOS}{fully overlapping subspaces}
\newacronym{fods}{FODS}{fully overlapping desired subspaces}
\newacronym{pos}{PODS}{partially overlapping desired subspaces}
\newacronym{wola}{WOLA}{weighted overlap-add}
\newacronym{rir}{RIR}{room impulse response}

\showchangesfalse

\makeatletter
\def\binew#1{\expandafter\gdef\csname ti@newref@#1\endcsname{}}
\binew{musluogluUnifiedAlgorithmicFramework2023}
\binew{musluoglu_distributed_2023}
\binew{dijkstra1959note}
\binew{scheiblerPyroomacousticsPythonPackage2018}
\binew{veaux2017vctk}
\binew{serizel_low-rank_2014}
\binew{taalAlgorithmIntelligibilityPrediction2011}
\AtBeginDocument{
  \let\ti@orig@bibitem\bibitem
  \renewcommand{\bibitem}[1]{%
    \normalcolor                       
    \@ifundefined{ti@newref@#1}{}{\ifshowchanges\color{blue}\fi}
    \ti@orig@bibitem{#1}%
  }%
}
\makeatother

\DeclareUnicodeCharacter{2212}{−}
\usepgfplotslibrary{groupplots,dateplot}
\usetikzlibrary{patterns,shapes.arrows,external}
\pgfplotsset{compat=newest}

\def\OJlogo{\vspace{-4pt}$<$Society logo(s) and publication title will appear here.$>$}
\def\seclogo{\vspace{10pt}$<$Society logo(s) and publication title will appear here.$>$}

\def\authorrefmark#1{\ensuremath{^{\textbf{#1}}}}

\begin{document}
\receiveddate{XX Month, XXXX}
\reviseddate{XX Month, XXXX}
\accepteddate{XX Month, XXXX}
\publisheddate{XX Month, XXXX}
\currentdate{XX Month, XXXX}
\doiinfo{XXXX.2022.1234567}

\markboth{}{P. Didier {et al.}: Distributed Multichannel Wiener Filtering for Topology-Unconstrained Wireless Acoustic Sensor Networks}

\title{Distributed Multichannel Wiener Filtering for Topology-Unconstrained Wireless Acoustic Sensor Networks}

\author{Paul Didier\authorrefmark{1}, Pourya Behmandpoor\authorrefmark{4}, Henri Gode\authorrefmark{2}, Toon van Waterschoot\authorrefmark{1}, Simon Doclo\authorrefmark{2,3}, Jörg Bitzer\authorrefmark{3}, and Marc Moonen\authorrefmark{1}}
\affil{STADIUS Center for Dynamical Systems, Signal Processing, and Data Analytics, Electrical Engineering Department (ESAT), KU Leuven, Leuven, Belgium}
\affil{Signal Processing Group, Department of Medical Physics, and with the Acoustics and Cluster of Excellence Hearing4all, Carl von Ossietzky Universit{\"a}t Oldenburg, Oldenburg, Germany}
\affil{Fraunhofer IDMT, Project Group Hearing, Speech and Audio Technology, Oldenburg, Germany}
\affil{Department of Electronics and Informatics, Vrije Universiteit Brussel, B-1050 Brussels, Belgium}
\corresp{Corresponding author: Paul Didier (email: phmdidier@proton.me).}
\authornote{This research work was carried out in the frame of Research Council KU Leuven project C14-21-0075 "A holistic approach to the design of integrated and distributed digital signal processing algorithms for audio and speech communication devices" and the European Union's Horizon 2020 research and innovation programme under the Marie Skłodowska-Curie Grant Agreement No. 956369: ``Service-Oriented Ubiquitous Network-Driven Sound — SOUNDS''. This paper reflects only the authors' views and the Union is not liable for any use that may be made of the contained information. The scientific responsibility is assumed by the authors.}

\begin{abstract}
This paper introduces the topology-independent distributed multichannel Wiener filter (TI-dMWF), a novel algorithm for distributed node-specific signal estimation in wireless acoustic sensor networks (WASNs) with unconstrained topologies. The TI-dMWF enables each node in the network to compute its centralized multichannel Wiener filter solution by exchanging only low-dimensional fused signals, without requiring iterative estimation, unlike state-of-the-art approaches such as the topology-independent distributed adaptive node-specific signal estimation (TI-DANSE) algorithm. The TI-dMWF is proven optimal when each source is observed by either all nodes or only one node. Theoretical analysis and numerical simulations confirm that it achieves centralized estimation performance in a single run. \change{Its latency as a function of the pruned-tree depth and its computational complexity are also analyzed. Its robustness is assessed in reverberant-room simulations under estimated second-order statistics, various network topologies, and deviations from the assumed observability model.}
\end{abstract}

\begin{IEEEkeywords}
Distributed signal processing, ad-hoc wireless acoustic sensor networks, multichannel Wiener filter, dimensionality reduction,
distributed noise reduction
\end{IEEEkeywords}

\glsunset{dansep}

\maketitle

\section{Introduction}\label{sec:intro}

\noindent

\IEEEPARstart{I}{n} recent years, devices capable of recording, processing, and exchanging audio data wirelessly have become ubiquitous, enabling the creation of \glspl*{wasn}, which have gained attention for applications such as environmental monitoring and smart homes~\cite{bertrand_applications_2011,natarajanComprehensiveReviewBeamformingBased2023}. Within a \gls*{wasn}, devices (nodes) equipped with one or more microphones (sensors), such as smartphones, laptops, or hearing aids, collaborate to perform audio signal processing tasks, e.g., noise reduction~\cite{doclo_reduced-bandwidth_2009,bertrand_distributed_2010,hassani_gevd-based_2016}, dereverberation~\cite{lohmannDereverberationAcousticSensor2023,changDistributedKalmanFiltering2023}, or sound source localization~\cite{griffinLocalizingMultipleAudio2015, cobos_survey_2017}. Compared to traditional fixed microphone arrays, \glspl*{wasn} offer greater spatial coverage, deployment flexibility, and robustness to node failures~\cite{bertrand_applications_2011}, but operate under constraints such as communication cost and potentially dynamic topologies, making the design of distributed algorithms that respect these constraints a key challenge.

\change{
This paper addresses node-specific signal estimation in a \gls*{wasn}, where each node estimates a signal of interest (e.g., speech) from its own sensor signals and signals sent by other nodes. The optimal centralized \gls*{lmmse} solution to this problem is provided as the \gls*{mwf}~\cite[Ch.2]{benestySpeechEnhancement2006}, but computing the \gls*{mwf} directly requires access to all sensor signals in the network, which is often infeasible due to communication cost constraints and in \glspl*{wasn} that are not \gls*{fc}.}
The \gls*{dmwf}~\cite{didier2025dmwf} lets each node of a \gls*{fc} \gls*{wasn} solve its node-specific \gls*{lmmse} problem distributively by exchanging only dimensionally reduced (fused) signals, addressing the communication cost constraint. However, topologies may not be \gls*{fc} and may vary over time. We here propose the \gls*{tidmwf}, which lets nodes of a topology-unconstrained \gls*{wasn} compute the centralized \gls*{mwf} as if they had access to all network signals, while exchanging only fused signals. \change{Here, \textit{topology-unconstrained} is a property of the \gls*{wasn} and \textit{\gls*{ti}} is a property of an algorithm: the \gls*{tidmwf} is a topology-independent algorithm over topology-unconstrained \glspl*{wasn}.} As the \gls*{dmwf}, it is iterationless, unlike the \gls*{tidanse}~\cite{szurley_topology-independent_2017} and \gls*{dansep}~\cite{didier2025fastconvergingdistributedsignalestimation} algorithms.

\change{The \gls*{tidanse} algorithm belongs to the \gls*{dasf} framework~\cite{musluogluUnifiedAlgorithmicFramework2023}, which the \gls*{dansf} framework~\cite{musluoglu_distributed_2023} extends to node-specific problems. Note that \gls*{dansep} is closely related to \gls*{dansf}~\cite{didier2025fastconvergingdistributedsignalestimation}. Unlike all these iterative algorithms, the \gls*{tidmwf} reaches the centralized \gls*{mwf} in a single discovery-estimation pass in \gls*{gls} scenarios (defined hereafter).}

Unlike the \gls*{dmwf}, which handles any \gls*{pos} scenario~\cite{didier2025dmwf}, the \gls*{tidmwf} targets \gls*{gls} scenarios, where each (speech or noise) source is observed by either all nodes (a \textit{global} source) or only one node (a \textit{local} source). Such scenarios arise, e.g., in airports or train stations: a global source is a signal propagating throughout the environment and picked up by all nodes (e.g., a public-address or alarm announcement), while a local source is observed by a single node (e.g., a nearby talker or localized noise). \change{The more general case where a source is observed by an arbitrary subset of nodes is described by the \gls*{pos} model; the \gls*{gls} model considered here remains a relevant special case whenever network-wide and strictly local sources are clearly separated.}
The \gls*{gls} assumption differs from the \gls*{fods} assumption~\cite{didier2025dmwf} (all speech sources observed by all nodes) needed for \gls*{tidanse} and \gls*{dansep} convergence and optimality~\cite{szurley_topology-independent_2017,didier2025fastconvergingdistributedsignalestimation}.

\section{Problem Statement}\label{sec:problem_statement}

We consider a \gls*{wasn} consisting of $K$ nodes, where node $k$ has $M_k$ sensors ($k\in\mathcal{K}:=\{1,...,K\}$) such that the total number of sensors is $M=\sum_{k\in\mathcal{K}}M_k$. Signals are complex-valued, e.g., representing processing in one filter-bank bin. The acoustic scenario includes ${\Qd}$ desired sources whose latent signals at time $t$ are grouped in $\mathbf{s}^\mathrm{lat}[t]\in\mathbb{C}^{{\Qd}\times1}$, and ${\Qn}$ noise sources with latent signals $\mathbf{n}^\mathrm{lat}[t]\in\mathbb{C}^{{\Qn}\times1}$, for a total of $Q = \Qd + \Qn$ sources. The sensor signals at all nodes are stacked in $\mathbf{y}[t]\in\mathbb{C}^{M\times 1}$. Omitting the time-index $[t]$ for conciseness, $\yk[]$ can be written as $\yk[] = \begin{bmatrix}
    \yk[1]^\T, \dots, \yk[K]^\T
\end{bmatrix}^\T = \sk[] + \nk[] + \vk[]$, where $\sk[]=\Ak[]\slat$ and $\nk[]=\Bk[]\nlat$ are the speech and noise components, with $\Ak[]\in\C[M][{\Qd}]$ and $\Bk[]\in\C[M][{\Qn}]$ the corresponding steering \change{matrices}, and $\vk[]\in\C[M]$ the self-noise.
In \gls*{gls} scenarios, this can be further expanded as:

\vspace{-1em}
\begin{equation}\label{eq:tidmwf:centr_signalmodel_expanded}
  \yk[] = \csk[] + \usk[] + \cnk[] + \unk[] + \vk[],
\end{equation}

\noindent
where $\sk[]=\csk[]+\usk[]$ and $\nk[]=\cnk[]+\unk[]$ decompose into global-source components $\csk[] = \cAk[]\slat$, $\cnk[] = \cBk[]\nlat$ and local-source components $\usk[] = \uAk[]\slat$, $\unk[] = \uBk[]\nlat$. The matrices $\cAk[]$, $\cBk[]$ (resp.\ $\uAk[]$, $\uBk[]$) equal $\Ak[]$, $\Bk[]$ with the local-source (resp.\ global-source) columns set to zero. As in the \gls*{dmwf}~\cite{didier2025dmwf}, the matrix obtained by removing the all-zeros columns from $[\cAk[]\:|\:\cBk[]]$ is assumed full column rank; its number of columns $\cQ$ is the \textit{total number of global sources}, and $\uQ = Q-\cQ$ the number of local sources. \change{A root node $k$ is assumed to satisfy $M_k\geq\cQ$, so that its upstream signal in \secref{sec:algorithm} (cf.~\eqref{eq:tidmwf:yuk}) is well-defined; a node with $M_k<\cQ$ may still participate as a non-root node (cf.~\secref{sec:tidmwf_practical}). Relaxing this for root nodes is left for future work (cf.~\secref{sec:conclusion}).}

Without loss of generality (w.l.o.g.), the desired signal at any node $k$ contains the contribution to the first $D$ sensors of node $k$ of all speech sources observed by node $k$ (global and local speech sources), i.e., $\dk = \Ek^\T\sk[]\in\C[D]\fa k\in\K$, where $\Ek\in\{0,1\}^{M\times D}$ extracts the first $D$ channels of $\sk$ from $\sk[]$. Also w.l.o.g., $D$ is assumed to be the same for all nodes.
To optimally estimate $\dk$ based on $\yk[]$ in the \gls*{lmmse} sense in a centralized setting, the centralized \gls*{mwf} can be computed at the fusion center as:

\vspace{-.5em}
\begin{align}\label{eq:tidmwf:centralized_problem}
  \hWk &= \argmin[{\mathbf{W}\in\C[M][D]}]
  \El[{
    \dk - \mathbf{W}^\Her\yk[]
  }] = \Ryy^{-1}\Rydk,
\end{align}

\noindent
where $\Ryy = \E[{\yk[]\yk[]^\Her}]$ and $\Rydk=\E[{
  \yk[]\dk^\Her
}]$ are \glspl*{scm}. To estimate these \glspl*{scm} in practical speech enhancement scenarios time-averaging strategies can be applied, 
\change{
  assuming ergodic signals.
  The first \gls*{scm} $\Ryy$ can be estimated during speech-active periods using, e.g., exponential averaging over incoming time-frames of $\yk[]$~\cite{bertrand_distributed_2010}. The second \gls*{scm} $\Rydk$ cannot be estimated directly as it involves the unavailable desired signal $\dk$. Instead, since $\Rydk = \Rss\Ek$ with $\Rss = \E[{\sk[]\sk[]^\Her}]$, and assuming wide-sense stationary noise, the on-off nature of speech is leveraged through a \gls*{vad} or \gls*{spp}~\cite{bahariDistributedMultiSpeakerVoice2017a, zhaoModelbasedDistributedNode2020, soudenGaussianModelBasedMultichannel2010a, tammenDNNBasedSpeechPresence2020} to estimate the noise-only \gls*{scm} $\Rnn$ in speech-absent frames, giving $\Rydk = (\Ryy - \Rnn)\Ek$. A \gls*{gevd}-based approach can alternatively be used~\cite{hassani_gevd-based_2016}. In practice, we may use a forgetting factor $0\ll\beta<1$ to recursively estimate $\Ryy$ and $\Rnn$ as:
}

\vspace{-.5em}
\begin{equation}\label{eq:expavg}
  \begin{split}
    \text{if VAD=1:}\:& \Ryy[t] = \beta\Ryy[t-1] + (1-\beta)\yk[k][t]\yk[k][t]^\Her,\\
    \text{if VAD=0:}\:& \Rnn[t] = \beta\Rnn[t-1] + (1-\beta)\yk[k][t]\yk[k][t]^\Her,
  \end{split}
\end{equation}

\noindent
where $\Ryy[t]$ and $\Rnn[t]$ denote \gls*{scm} at time-frame $t$.

\section{Topology-Independent Distributed MWF (TI-dMWF)}\label{sec:algorithm}

Without a fusion center, communication cost constraints and a possibly non-\gls*{fc}, time-varying topology inhibit the direct computation of the centralized \gls*{mwf}~\eqref{eq:tidmwf:centralized_problem} at each node $k$. The \gls*{tidmwf} described in this section lets each node of any topology-unconstrained \gls*{wasn} compute the centralized \gls*{mwf}~\eqref{eq:tidmwf:centralized_problem} distributively by exchanging only dimensionally reduced (fused) sensor signals.
The \gls*{tidmwf} is defined with respect to one specific node $k$ and only allows the estimation of the desired signal $\dk$ at this node. The \gls*{tidmwf} steps must thus be repeated for each node in the \gls*{wasn}.

The desired signal estimate at a particular node $k$ is obtained through cascaded estimation-fusion processes. To enable this cascade of operations, the \gls*{wasn} is first pruned to a tree.
For instance, \change{a} \gls*{mst} can be computed from any \gls*{wasn} topology using Prim's algorithm~\cite{prim1957shortest}, then oriented by choosing a root node. In the resulting rooted tree, a non-root node with a single neighbor is a leaf node and any other non-root node is an internal node; each non-root node has one downstream neighbor (towards the root) and each internal node at least one upstream neighbor (towards the leaves). We let $\Uk[q]$ denote the upstream neighbors of node $q$ and $\Ukb[q]$ all nodes upstream of $q$, with $\Uk[q]\subseteq\Ukb[q]$.

Consider a node $k$ and let it be the root of the tree after pruning.
As in the \gls*{dmwf}~\cite{didier2025dmwf}, we seek a set of signals, stacked into an observation vector $\ty$ at node $k$, from which $\dk$ can be estimated.
To this end, a sum-and-send data flow from the leaf nodes to the root node $k$ is defined. To outline it, we focus on one non-root node $q\in\Kbq[k]= \K\backslash\{k\}$.
In the considered \gls*{gls} scenario, the \gls*{tidmwf} fused signal dimension is $\cQ$, the total number of \textit{global} sources, so that the fused signal encompasses all accessible information about the global-sources subspace (recalling that all sources are mutually uncorrelated).
\change{If the total dimension $\hMk[q]$ of the signals available at node $q$ (defined in~\eqref{eq:tidmwf:hy}) does not exceed $\cQ$, i.e., $\hMk[q]\leq \cQ$, fusing them into a $\cQ$-dimensional signal would not reduce their dimensionality. Node $q$ then directly forwards its available signals to its downstream neighbor, bypassing the fusion operation. This \textit{mixed-network} case is discussed in~\secref{sec:tidmwf_practical}. For exposition, we assume $\hMk[q]> \cQ\fa q\in\Kbq[k]$ in the remainder of this section.}

Aiming at estimation of $\dk$ at root node $k$, node $q$ will send a fused signal $\hzktq[q][k]$ to its (only) downstream neighbor in the tree topology. This signal is recursively defined as a linear combination of the local sensor signals $\yk[q]$ with the signals received by node $q$ from its own upstream neighbors (if $q$ is a leaf node, then $\hyqtk$ below is simply $\yk[q]$), i.e.:

\vspace{-.5em}
\begin{align}
  \forall\:q\in\Kbq[k]:\hzktq[q][k] &=
  \hPktq[q][k]^\Her\hyqtk\in\C[{\hQk[kq]}],\label{eq:tidmwf:hzktq}\\
  \with
  \hyqtk &=
  \begin{bmatrix}
    \yk[q]^\T,\hzktq[u_1][k]^\T,\dots,\hzktq[u_B][k]^\T
  \end{bmatrix}^\T\in\C[{\hMk[q]}],\label{eq:tidmwf:hy}\\
  \where
  \hMk[q] &=
  M_q + |\Uk[q]|\hQk[ku_i],
\end{align}

\noindent
where $\{u_1,\dots,u_B\}$ are the indices of the upstream neighbors of node $q$ (with $B=|\Uk[q]|$), and $\hPktq[q][k]\in\C[{\hMk[q]}][{\hQk[kq]}]$ is the fusion matrix at node $q$ directed towards root node $k$.
For notation simplicity, we omit to refer to the root node $k$ in $\hzktq[q]$, $\hPktq[q]$, and $\hy[q]$ but it is understood that these quantities are defined with respect to a particular root node $k$. When a tree with another root is pruned, $B=|\Uk[q]|$ and $\hy[q]$ change (and so do $\hPktq[q]$ and $\hzktq[q]$) depending on the new set $\{\Uk[q]\}_{q\in\K}$.
\change{
It is noted that, for any leaf node $l$ (i.e., any non-root node with a single neighbor), $\hy[l] = \yk[l]$ and thus $\hMk[l] = M_l$.}

The \gls*{tidmwf} is composed of two distinct steps. When aiming at a particular root node $k$, the first step is the \textit{discovery step}, where the fusion matrix $\hPktq[q][k]$ and fused signal $\hzktq[q][k]$ are computed at each node $q\in\Kbq[k]$. This discovery step can itself be subdivided into two parts: (i) an \textit{upstream data flow}, where fused signals are transmitted from the root node $k$ to the leaf nodes, and (ii) a \textit{downstream data flow}, where fused signals are transmitted from the leaf nodes to the root node $k$. The second step is an \textit{estimation step} where the root node estimates $\dk$ using the observation vector $\hy$. 

\subsection{Discovery Step: Upstream Data Flow}\label{sec:tidmwf_upstream_data_flow}

In the upstream data flow, the root node $k$ floods a signal $\yuk[q]$ through the tree, such that every other node in $\Kbq[k]$ eventually has access to $\yuk[q]$.
This signal is defined as a subset of $\yk$:

\vspace{-.5em}
\begin{align}
  \yuk[q] &= \Euk[q]^\T\yk\in\C[{\hQk[kq]}],\label{eq:tidmwf:yuk}
\end{align}
  
\noindent
where $\Euk[q]= [\mathbf{I}_{\hQk[kq]}\:|\:\zer]^\T\in\{0,1\}^{M_k\times \hQk[kq]}$ is a selection matrix.
\change{For simplicity,~\eqref{eq:tidmwf:yuk} assumes all $\cQ$ global sources are active and that $\cQ$ is known a priori or estimated by node $k$ via source enumeration~\cite{williamsUsingSphericityTest1990,brcichDetectionSourcesUsing2002,stoicaModelorderSelectionReview2004,waxDetectionNumberSignals2021}.}

\subsection{Discovery Step: Downstream Data Flow}\label{sec:tidmwf_downstream_data_flow}

\change{
The downstream data flow runs from the leaf nodes towards the root node $k$. Each node $q\in\Kbq[k]$ receives the fused signals $\{\hzktq[u][k]\}_{u\in\Uk[q]}$ from its upstream neighbors,
} builds $\hy[q]$ via~\eqref{eq:tidmwf:hy}, and uses it with $\yuk[q]$ to compute its fusion matrix (towards root $k$) by solving the \gls*{lmmse} problem:

\vspace{-.5em}
\begin{align}\label{eq:tidmwf:Pktq_alt_ti}
  \hPktq[q][k] &= \argmin[{\mathbf{P}_q\in\C[{\hMk[q]}][{\hQk[kq]}]}]
  \El[{
    \yuk[q] - \mathbf{P}_q^\Her\hyqtk[q]
  }],
\end{align}
\vspace{-.5em}
\begin{align}\label{eq:tidmwf:Pktq_alt_ti_mwf_2}
  \Leftrightarrow \hPktq[q][k] = \Rhyhy[q][k]^{-1}\Rhyqyuq[q]\in\C[{\hMk[q]}][{\hQk[kq]}]\fa q\in\Kbq[k],
\end{align}

\noindent
where $\Rhyhy[q][k] = \E[{
  \hyqtk[q]\hyqtk[q]^\Her
}]\in\C[{\hMk[q]}][{\hMk[q]}]$ and $\Rhyqyuq[q] = \E[{
  \hyqtk[q]\yuk[q]^\Her
}]\in\C[{\hMk[q]}][{\hQk[kq]}]$. 
Node $q$ can then compute $\hzktq[q][k]$ via~\eqref{eq:tidmwf:hzktq} and send it to its own downstream neighbor. \change{For a leaf node $l$,~\eqref{eq:tidmwf:Pktq_alt_ti} and~\eqref{eq:tidmwf:Pktq_alt_ti_mwf_2} also hold, with $\hy[l] = \yk[l]$ and $\hMk[l] = M_l$.}

When the tree-pruned \gls*{wasn} is a star (every node $q\in\Kbq[k]$ directly connected to the root $k$), the \gls*{tidmwf} reduces to the \gls*{dmwf}~\cite{didier2025dmwf}: the upstream flow sends $\yuk[q]$ (equating $\bykq$ in~\cite{didier2025dmwf}) to every $q$, and the downstream flow returns $\hzktq[q][k]$ (equating $\zktq[q][k]$) to $k$.

\subsection{Estimation Step}\label{sec:tidmwf_estimation}

At the end of the downstream data flow, root node $k$ has access to $\{\hzktq[q][k]\}_{q\in\Uk}$ and can build its own observation vector $\hy$ as in~\eqref{eq:tidmwf:hy}. Node $k$ can then estimate its desired signal based on $\hy$ via:

\vspace{-.5em}
\begin{align}\label{eq:tidmwf:estimation_lmmse_ti}
  \tWTI &= \argmin[{\mathbf{W}_k\in\C[{\hMk}][D]}]
  \El[{
    \dk - \mathbf{W}_k^\Her\hy
  }],
\end{align}
\vspace{-.5em}
\begin{align}\label{eq:tidmwf:tidmwf_mwf}
  \Leftrightarrow \tWTI = \Rhyhy[k][]^{-1}\Rhydk[k][],
\end{align}

\noindent
where $\Rhydk[k][]=\E[{
  \hy\dk^\Her
}]$.
\change{In a similar way as for the centralized \gls*{mwf} from~\eqref{eq:tidmwf:centralized_problem}, the two \glspl*{scm} necessary to compute~\eqref{eq:tidmwf:tidmwf_mwf} may be estimated based on a \gls*{vad} (or an \gls*{spp}) and time-averaging as in~\eqref{eq:expavg}.}
The \gls*{tidmwf} desired signal estimate is then obtained as $\dhatk[k,\text{\gls*{tidmwf}}] = (\tWTI)^\Her\hy$. All three steps are repeated for each node $k\in\K$ to estimate $\{\dk\}_{k\in\K}$. When the root index $k$ changes, so do $\hy[q]$ (and hence $\hPktq[q]$ and $\hzktq[q]$), through the new sets $\{\Uk[q]\}_{q\in\Kbq[k]}$.

\subsection{TI-dMWF Overview and Optimality}\label{sec:tidmwf_optimality}

Each node $k$ performs the estimation step at every time-frame, which requires the fused signals $\{\hzktq[q][k]\}_{q\in\Uk}$ from the downstream data flow towards $k$. These steps are thus performed every frame. The fusion-matrix computation and the upstream data flow, however, may run less frequently (e.g., every $N_\mathrm{us}> 1$ frames).
The \gls*{tidmwf} is summarized in~\algref{alg:tidmwf}.

\noindent
\textit{Theorem 1 (\gls*{tidmwf} optimality):} 
Under the signal model~\eqref{eq:tidmwf:centr_signalmodel_expanded} (\gls*{gls} scenario), the \gls*{tidmwf} desired signal estimate is equal to the centralized desired signal estimate at node $k\fa k\in\K$, when $\hPktq[q][k]$ is defined as in~\eqref{eq:tidmwf:Pktq_alt_ti_mwf_2}$\fa q\in\Kbq[k]$, i.e., $(\tWTI)^\Her\hy = \hWk^\Her\yk[]\fa k\in\K$ where $\hWk$ is defined in~\eqref{eq:tidmwf:centralized_problem} and $\tWTI$ is defined in~\eqref{eq:tidmwf:tidmwf_mwf}. \textit{Proof}: see Appendix A.
\vspace{.5em}\newline
\change{
\noindent
\textit{Remark 1 (Failure under \gls*{pos} scenarios):} The proof of Theorem~1 relies on the inter-node uncorrelatedness of the non-global components $\{\uyk[q]\}_{q\in\K}$, which holds under the \gls*{gls} model (every source is global or observed by a single node). It is used twice: the block-diagonality of $\Gam$ in~\eqref{eq:proof_solution_twonodes} admits a factorization of the joint \gls*{lmmse} filter $\tW[l'l]$ into locally computable $\cQ$-dimensional fusion matrices, and it justifies replacing the unobservable $\Euk^\T\cyk$ by the measurable $\Euk^\T\yk$ at every non-root node. Under a \gls*{pos} model, a source observed by a strict subset of two or more nodes makes $\E[{\uyk[q]\uyk[q']^\Her}]\neq\zer$ for some $q\neq q'$, so both steps fail: the fused signals are no longer a sufficient statistic and the algorithm does not reach the centralized \gls*{mwf}. This suboptimality is quantified in \secref{ssec:robustness}.
}

\begin{algorithm}[!ht]
  \caption{The TI-dMWF.}\label{alg:tidmwf}
  \begin{algorithmic}[1]
    \small
    \FOR{time-frame index $t=0,1,\dots$}
      \FOR{$k\in\K$}
        \STATE The \gls*{wasn} is pruned to a tree with root $k$.
        \IF{{
          $\mod(t,N_\mathrm{us})=0$}}
          \STATE $\yuk[k][t]$ is flooded from $k$ through the tree.
        \ENDIF
        \FOR{$q$ from leaves to root $k$ (excluded)}
          \STATE Node $q$ gathers $\{\hzktq[q'][k][t]\}_{q'\in\Uk[q]}$ and builds $\hy[q][t]$ as in~\eqref{eq:tidmwf:hy}.
          \STATE Node $q$ estimates $\Rhyhy[q][k]$ using $\hy[q][t]$ and $\Rhyqyuq[q]$ using $\yuk[q][t]$ (or $\yuk[q][t']$ where $t' {<}t$ if $\mod(t,N_\mathrm{us}){\neq}0$).
          \STATE Node $q$ computes $\hPktq[q][k]$ as in~\eqref{eq:tidmwf:Pktq_alt_ti_mwf_2}.
          \STATE Node $q$ computes $\hzktq[q][k][t]$ as in~\eqref{eq:tidmwf:hzktq} and sends it downstream.
        \ENDFOR
        \STATE Node $k$ gathers $\{\hzktq[q][k][t]\}_{q\in\Uk}$ and builds $\hy[k][t]$ as in~\eqref{eq:tidmwf:hy}.
        \STATE Node $k$ estimates $\Rhyhy[k][]$ and $\Rhydk[k]$.
        \STATE Node $k$ computes $\tWTI[k]$ as in~\eqref{eq:tidmwf:tidmwf_mwf} then $\dhatk[k,\text{\gls*{tidmwf}}][t]$.
      \ENDFOR
    \ENDFOR
  \end{algorithmic}
\end{algorithm}

\subsection{Practical Considerations}\label{sec:tidmwf_practical}

\change{Since the pruned tree (and hence the fusion matrices) depends on the root, each non-root node $q$ generally maintains a distinct fusion matrix $\hPktq[q][k]$ for each possible root $k\in\Kbq[q]$, i.e., $K-1$ in total, with the tree (re-)oriented per root.

When the topology changes (e.g., due to a link failure or a node joining/leaving), the pruned tree is recomputed for each affected root, and the fusion matrices $\hPktq[q][k]$ re-estimated only at nodes $q$ whose upstream set $\Uk[q]$ (hence $\hyqtk[q]$) changes. The matrices of unaffected nodes remain valid. We advocate a \gls*{spt}~\cite{dijkstra1959note} rooted at each $k$: since the downstream flow runs every frame while the upstream flow runs every $N_\mathrm{us}$ frames, the dominant per-frame latency grows with the tree \textit{depth}, which an \gls*{spt} minimizes.

The flooding of $\yuk[q]$ need not be synchronous: since the upstream flow is decoupled from the per-frame downstream flow, node $q$ may compute $\hPktq[q][k]$ from a delayed $\yuk[q][t']$ ($t'<t$), as the \gls*{scm} estimation relies on time-averaged statistics (cf.~\eqref{eq:expavg}) that vary slowly relative to the frame rate.

The \gls*{tidmwf} also accommodates \textit{mixed networks} where only some nodes apply a fusion operation. A node $q$ for which fusion would not reduce the forwarded dimension ($\hMk[q]\leq\cQ$) instead forwards $\hyqtk[q]$ unfused, which is then treated by its downstream neighbor as additional channels in~\eqref{eq:tidmwf:hy}. Theorem~1 optimality is preserved, as forwarding unfused signals retains the global-sources subspace information and the cascaded-fusion argument of Appendix~A holds for any subset of fusing nodes.

The number of global sources $\cQ$, setting the dimension of $\yuk[q]$ in~\eqref{eq:tidmwf:yuk}, must in practice be estimated by source enumeration (cf.~\secref{sec:tidmwf_upstream_data_flow}). For a mis-estimated $\cQ'\neq\cQ$, fixed network-wide for a given root $k$, every downstream node targets the same $\cQ'$-dimensional signal in~\eqref{eq:tidmwf:Pktq_alt_ti}. Underestimation ($\cQ'<\cQ$) makes $\yuk[q]$ span only a $\cQ'$-dimensional subset of the global-sources subspace, losing information and yielding a suboptimal estimate. Overestimation ($\cQ'>\cQ$) preserves optimality at the cost of a higher communication cost ($\cQ'$-dimensional exchanges), provided $\Rhyhy[q][k]$ stays full-rank with the extra channels, which holds generically owing to per-sensor self-noise. This assumes $\cQ'\leq M_k$; the case $\cQ'>M_k$, for which~\eqref{eq:tidmwf:yuk} is ill-defined, is left for future work (cf.~\secref{sec:conclusion}).}

\section{{Communication Cost Analysis}}\label{sec:comm_bandwidth_usage}

We analyze and compare the communication cost of the \gls*{tidmwf} and \gls*{tidanse}, assuming peer-to-peer signal exchanges and a \gls*{tidmwf} upstream data flow every $N_\mathrm{us}\geq 1$ frames (cf.~\secref{sec:algorithm}).
Note that \gls*{dansep} has the same communication cost as \gls*{tidanse}~\cite{didier2025fastconvergingdistributedsignalestimation}.

\change{
For one updating node $k$, the upstream data flow exchanges $N_{\mathrm{up}} = (K-1)\hQk[kq]$ signal channels (propagating $\yuk[q]$ from root to leaves), and similarly for the downstream data flow, $N_{\mathrm{down}} = (K-1)\hQk[kq]$. As the downstream flow runs continuously, the per-frame total is between $N_{\mathrm{down}}$ and $N_\mathrm{down} + N_\mathrm{up}$, depending on the upstream-flow frequency.
Both flows are repeated for each node $k$. Some exchanges could be saved when nodes share upstream or downstream neighbors, but for simplicity we leave this optimization for future work.

For comparison, \gls*{tidanse} exchanges $N_{\text{TI-D}}=2(K-1)\Qd$ signal channels per iteration~\cite{szurley_topology-independent_2017}. While this is lower than the \gls*{tidmwf}, \gls*{tidanse} requires \textit{multiple iterations} to approach the centralized \gls*{mwf}. The single-pass nature of the \gls*{tidmwf} thus generally yields a lower \textit{cumulative} communication cost than the iterative baselines whenever near-centralized accuracy is required. This has been verified in numerical experiments, omitted here for conciseness.
}

\change{
\section{Latency and Complexity Analysis}\label{sec:latency_complexity}

Real-time feasibility depends on the per-frame latency $L_k$, dominated in the \gls*{tidmwf} by the downstream data flow at every frame. Let $\delta_{\mathrm{hop}}$ be the average per-hop delay (transmission and per-node time to compute~\eqref{eq:tidmwf:hzktq}). Since each node $q$ must gather its upstream neighbors' fused signals before transmitting its own fused signals, the per-frame latency for root $k$ is $L_k = \Delta_k \delta_{\mathrm{hop}}$, with $\Delta_k$ the pruned-tree depth.

Topology-independence lets one choose a pruning strategy minimizing this depth: whereas an \gls*{mst} minimizes total communication power but may yield deeper trees, an \gls*{spt} rooted at $k$ minimizes $\Delta_k$~\cite{dijkstra1959note}. The real-time constraint $L_k < T_{\mathrm{shift}}$ (with $T_{\mathrm{shift}}$ the STFT frame shift) requires $\delta_{\mathrm{hop}} < T_{\mathrm{shift}}/\Delta_k$; for $T_{\mathrm{shift}} = 20$~ms and $\Delta_k=4$, this allows $\delta_{\mathrm{hop}}$ up to $5$~ms, which is attainable by low-latency wireless protocols.

A separate \textit{discovery latency} is associated with the upstream signal flooding. Since the discovery step runs only every $N_{\mathrm{us}}$ frames, independently of the per-frame estimation path, the upstream statistics may be delivered asynchronously or with delay without affecting optimality in slowly varying environments. Moreover, once the \glspl*{scm} are well-estimated, the \gls*{tidmwf} reaches centralized performance in a single discovery cycle.

The per-frame computational complexity of the \gls*{tidmwf} stays low by decoupling the frequent downstream fusion from the infrequent upstream discovery. In the \textit{estimation step}, each node performs a single matrix-vector product~\eqref{eq:tidmwf:hzktq}, of complexity $O(\cQ \hMk[q])$, plus per-frame rank-1 \gls*{scm} updates~\eqref{eq:expavg} costing $O(\hMk[q]^2 + \cQ\hMk[q])$. The most intensive task, solving~\eqref{eq:tidmwf:Pktq_alt_ti} for $\hPktq[q][k]$ at $O(\hMk[q]^3)$, runs only in the \textit{discovery step}, every $N_{\mathrm{us}}$ frames, giving a mean per-frame complexity $O(\cQ \hMk[q] + \hMk[q]^2 + \hMk[q]^3/N_{\mathrm{us}})$. In contrast, \gls*{tidanse} has a cheaper per-iteration update (an \gls*{lmmse} problem of size $M_k + \cQ$ at the updating node) but needs many such updates across frames. Finally, a topology change costs only tree re-pruning ($O(K|E| \log K)$ per node via Dijkstra's algorithm~\cite{dijkstra1959note} for $K$ root-specific \glspl*{spt}) plus fusion-matrix re-estimation at the affected nodes.}

\section{Numerical Simulations}\label{sec:res}

\change{This section reports four experiments. The first (\secref{ssec:oracle_validation}) validates the framework using oracle \glspl*{scm} in idealized scenarios. The second (\secref{ssec:robustness}) also uses oracle \glspl*{scm}, isolating model mismatch from estimation errors to assess robustness to deviations from the \gls*{gls} assumption. The third and fourth (\secref{ssec:estscm}) evaluate the \gls*{tidmwf} in simulated reverberant scenarios with \glspl*{scm} estimated from the microphone signals, assessing respectively the convergence under estimated \glspl*{scm} and the influence of the tree-pruning strategy.}

\subsection{Validation under Oracle SCMs}\label{ssec:oracle_validation}

\change{As a first experiment, the framework is validated using oracle \glspl*{scm}.} The \gls*{tidmwf} is compared in an ideal setting to \gls*{tidanse}~\cite{szurley_topology-independent_2017} (with the filter-normalization of~\cite{didier2024tigevddanse}) and \gls*{dansep}~\cite{didier2025fastconvergingdistributedsignalestimation}, validating the \gls*{tidmwf} framework and assessing the convergence of the iterative baselines. \change{Since \gls*{dansep} is closely related to \gls*{dansf}~\cite{musluoglu_distributed_2023,didier2025fastconvergingdistributedsignalestimation}, this also indirectly contrasts the \gls*{tidmwf} with the node-specific extension of the \gls*{dasf} framework.}
The \glspl*{scm} at the updating node of \gls*{tidanse} and \gls*{dansep} may be singular in general \gls*{gls} scenarios, so the comparison is conducted only in \gls*{cgls} scenarios where all speech sources are observed by all nodes.

In the idealized setting, a topology-unconstrained \gls*{wasn} of $K=6$ nodes with $M_k=5\fa k\in\K$ is deployed with $\Qd=2$ speech and $\Qn=2$ noise sources (so $M_k\geq Q\fa k\in\K$).
The steering matrix entries used in the signal model~\eqref{eq:tidmwf:centr_signalmodel_expanded} are independently drawn from a standard normal distribution.
Each node $k\in\K$ aims to estimate its own single-channel ($D{=}1$) desired signal $d_k$.

All \glspl*{scm} needed for the \glspl*{mwf} in~\eqref{eq:tidmwf:centralized_problem},~\eqref{eq:tidmwf:Pktq_alt_ti_mwf_2}, and~\eqref{eq:tidmwf:tidmwf_mwf} (and cf.~\cite{szurley_topology-independent_2017,didier2025fastconvergingdistributedsignalestimation} for \gls*{tidanse} and \gls*{dansep}) are computed from oracle knowledge of the steering matrices, source powers, and self-noise power. The centralized \gls*{scm} is $\Ryy = \Rss + \Rnn + \Rvv= \Ak[]\Rsslat\Ak[]^\Her + \Bk[]\Rnnlat\Bk[]^\Her + \Rvv$, with $\Rsslat$, $\Rnnlat$ diagonal latent speech/noise power matrices and $\Rvv$ the (diagonal) self-noise \gls*{scm}.
\change{
The latent power is set to $1$ for every speech and noise source, and the self-noise power to $0.01$ for every sensor.}
The \glspl*{scm} required for the distributed algorithms are computed based on the centralized \glspl*{scm} and the fusion matrices. For instance, in the \gls*{tidmwf} estimation step, the \gls*{scm} $\Rhyhy[k][]$ in~\eqref{eq:tidmwf:tidmwf_mwf} is computed as $\Rhyhy[k][] = \E[{
    \hyqtk[k]\hyqtk[k]^\Her
  }] = \Dk^\Her\Ryy\Dk$, where $\Dk$ arranges the fusion matrices $\{\Pktq[q]\}_{q\in\Kbq[k]}$ (such that $\WkTI = \Dk\tWTI$, with $\WkTI$ the network-wide \gls*{tidmwf} filter at node $k$; see Appendix~A). The other \glspl*{scm} (for the \gls*{tidmwf} and for each \gls*{tidanse}/\gls*{dansep} iteration) follow analogously, omitted for brevity.

In each simulation, the \gls*{wasn} is generated as a non-fully connected graph with connectivity $C=0.5$ (as defined in~\cite{didier2025fastconvergingdistributedsignalestimation}).
This graph is pruned to a tree using the \gls*{mmut} strategy~\cite{didier2025fastconvergingdistributedsignalestimation}, a constrained variant of Kruskal's \gls*{mst} algorithm~\cite{kruskal1956shortest} shown to speed up \gls*{dansep} convergence relative to \gls*{mst} pruning.

Two evaluation metrics are used: the network-wide filter error $\mathrm{MSE}_{W} = \tfrac{1}{K}\sum_{k\in\K}\|\Wk-\hWk\|_\mathrm{F}^2$, measuring the difference between the distributed ($\Wk$) and centralized ($\hWk$) network-wide filter coefficients\footnote{For full expressions of the network-wide filters please refer to \change{\eqref{eq:proof_solution_twonodes}} for the \gls*{tidmwf},~\cite{szurley_topology-independent_2017} for \gls*{tidanse}, and~\cite{didier2025fastconvergingdistributedsignalestimation} for \gls*{dansep}.}, and the desired-signal error $\mathrm{MSE}_{d} = \tfrac{1}{K}\sum_{k\in\K}\|\dk-\dhatk\|_2^2$, quantifying the error in the desired-signal estimates.

\begin{figure*}[t]
  \centering
  \begin{subfigure}[b]{0.45\textwidth}
    \centering
    \includegraphics[width=\linewidth]{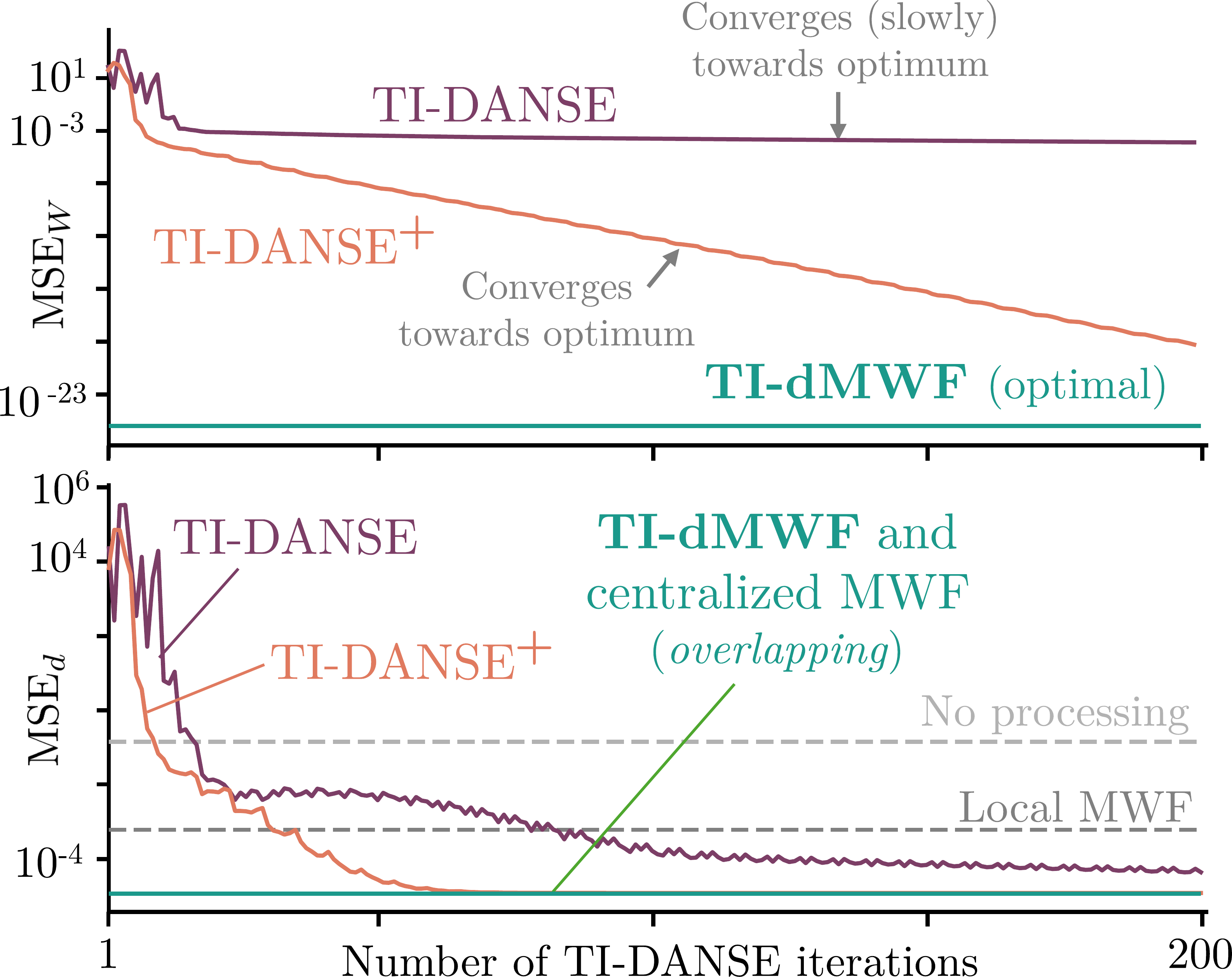}
    \caption{}\label{fig:tidmwf:tidmwf_res}
  \end{subfigure}\hfill
  \begin{subfigure}[b]{0.45\textwidth}
    \centering
    \includegraphics[width=\linewidth]{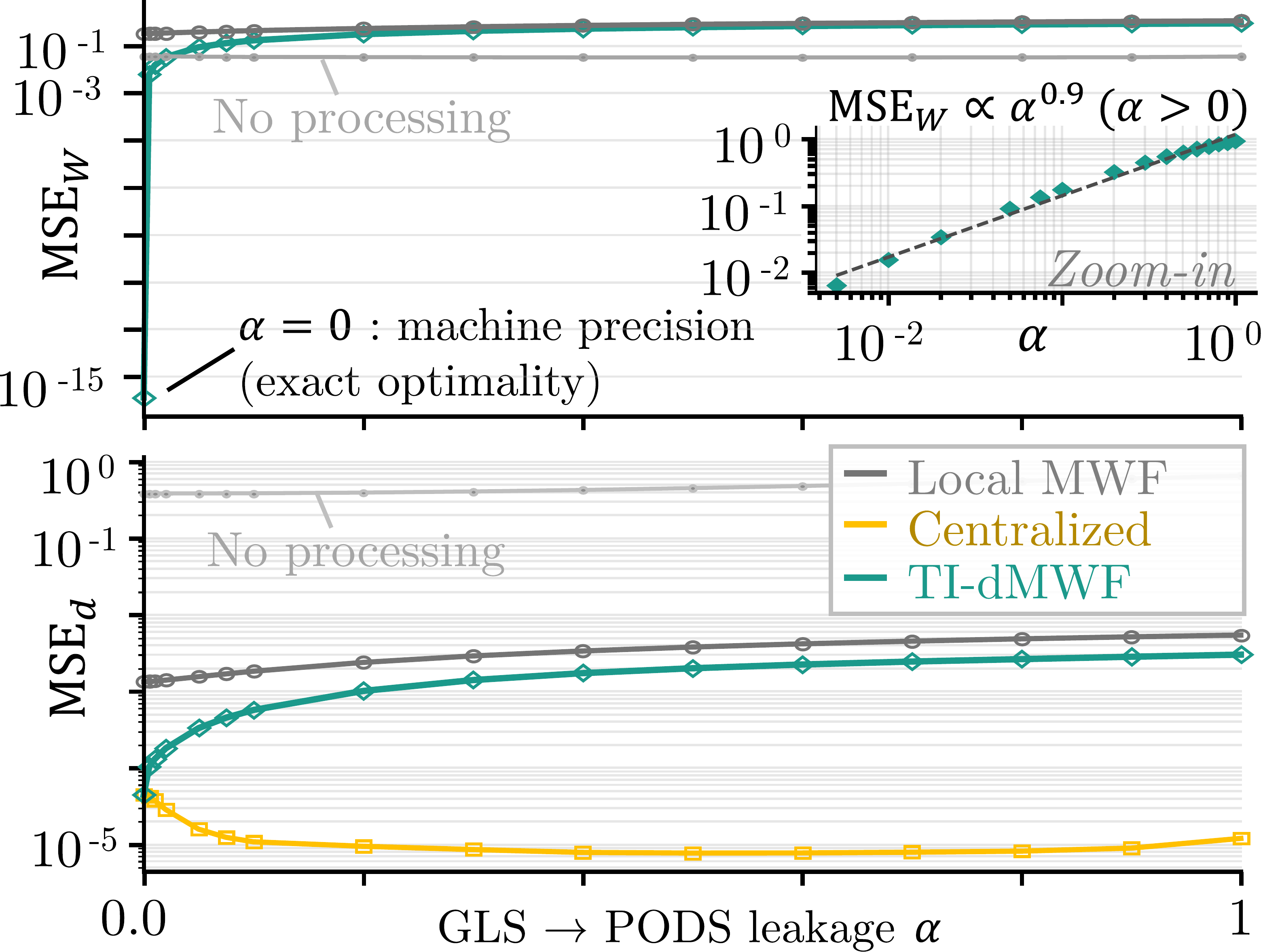}
    \caption{}\label{fig:tidmwf:glspods}
  \end{subfigure}
  \caption{\gls*{tidmwf} performance, $\mathrm{MSE}_{W}$ (top) and $\mathrm{MSE}_{d}$ (bottom). (a)~Versus iteration over $20$ \gls*{cgls} oracle scenarios ($C=0.5$), comparing the \gls*{tidmwf}, \gls*{tidanse}, and \gls*{dansep}. (b)~Versus the observability-leakage $\alpha$ over $20$ \gls*{gls} oracle scenarios, with local, centralized, and unprocessed references (log-log inset).}
  \label{fig:tidmwf:mse}
\end{figure*}

\figref{fig:tidmwf:tidmwf_res} compares the \gls*{tidmwf} to \gls*{tidanse} and \gls*{dansep} on both metrics. The $\mathrm{MSE}_d$ is also computed for $\dhatk$ set to (i) the unprocessed first sensor signal $y_{k,1}$ ($D{=}1$ here) and (ii) the local \gls*{mwf} estimate from $\yk$ alone. Results are averaged over 20 random \gls*{cgls} scenarios, using a geometric mean for $\mathrm{MSE}_{W}$ (for a clear logarithmic-axis depiction~\cite{didier2025fastconvergingdistributedsignalestimation}) and an arithmetic mean for $\mathrm{MSE}_{d}$.

\change{
To simulate observability patterns in \gls*{cgls} scenarios, $K\times Q$ \textit{observability matrices} (with entries $0$ for an unobserved and $1$ for an observed source) are randomly generated such that all speech sources are observed by all nodes and each noise source is observed by a single node or by all nodes.}

The results show that the \gls*{tidmwf} achieves the same performance as the centralized \gls*{mwf} and does so without relying on iterations, which is visible through the machine-precision $\mathrm{MSE}_W$ and the equal $\mathrm{MSE}_d$ between the centralized \gls*{mwf} and \gls*{tidmwf}. Conversely, \gls*{tidanse} requires many iterations to converge towards the centralized \gls*{mwf}. While \gls*{dansep} converges faster than \gls*{tidanse} (we use \gls*{mmut} pruning at connectivity $C=0.5$), it still requires many iterations to reach the centralized \gls*{mwf}. After about 60 iterations, its $\mathrm{MSE}_d$ nearly matches the centralized \gls*{mwf} even though its $\mathrm{MSE}_W$ has not reached that of the \gls*{tidmwf}, suggesting that a moderately low $\mathrm{MSE}_W$ can suffice in practice. To outperform the local \gls*{mwf}, \gls*{dansep} and \gls*{tidanse} require around 30 and 82 iterations, respectively, in this 6-node setup.

\subsection{Robustness to GLS Deviations}\label{ssec:robustness}

\change{The second experiment assesses \gls*{tidmwf}'s sensitivity to deviations from the \gls*{gls} assumption via a leakage parameter $\alpha\in[0,1]$. Sources unobserved under the assumed \gls*{gls} pattern leak into the corresponding microphone signals with gain $\alpha$, so $\alpha=0$ is the strict \gls*{gls} scenario and increasing $\alpha$ injects each source's energy into the nodes the pattern treats as non-observing. The algorithm keeps operating under the assumed pattern as the actual signals depart from it. To isolate model mismatch, it uses oracle \glspl*{scm} and the idealized setup of \secref{ssec:oracle_validation} with random steering matrices.}

\change{\figref{fig:tidmwf:glspods} shows the \gls*{tidmwf} $\mathrm{MSE}_W$ and $\mathrm{MSE}_d$ versus $\alpha$, with the centralized, local, and unprocessed references. At $\alpha=0$ (strict \gls*{gls}), the \gls*{tidmwf} attains the centralized \gls*{mwf} exactly, its network-wide filter coinciding with the centralized \gls*{mwf} to machine precision ($\mathrm{MSE}_W\approx10^{-16}$) and its estimate reaching the centralized $\mathrm{MSE}_d$, as predicted by Theorem~1. Any non-zero leakage immediately breaks this exact optimality, with no finite tolerance band: $\mathrm{MSE}_W$ leaves machine precision and rises continuously. This matches Remark~1 of \secref{sec:tidmwf_optimality}: once a source is observed by a strict subset of nodes, the non-global components $\{\uyk[q]\}_{q\in\K}$ become inter-node correlated, the block-diagonality of $\Gam$ in~\eqref{eq:proof_solution_twonodes} is lost, and the fused signals are no longer a sufficient statistic. Exact optimality is confined to the strict \gls*{gls} scenario, but away from it the \gls*{tidmwf} degrades smoothly and monotonically with $\alpha$. Nonetheless, across the whole $\alpha$ range the \gls*{tidmwf} network-wide filter stays closer to the centralized \gls*{mwf} than the local \gls*{mwf}, and its estimate stays more accurate than the local \gls*{mwf} and the unprocessed input.}

\subsection{Performance under Estimated SCMs}\label{ssec:estscm}

\change{Beyond the idealized setting of \secref{ssec:oracle_validation}, the \gls*{tidmwf} is evaluated in reverberant scenarios with all \glspl*{scm} estimated from the microphone signals. \glspl*{rir} are generated with the \gls*{ism} in Pyroomacoustics~\cite{scheiblerPyroomacousticsPythonPackage2018} for a $5\times 5\times 3$~m room with $T_{60}=0.2$~s.}

\change{As in \secref{ssec:oracle_validation}, the \gls*{wasn} has $K=6$ nodes with $M_k=5$ microphones (compact $3$~cm-radius arrays), placed at least $0.25$~m from the walls, with a $C=0.5$ ad-hoc topology pruned via the \gls*{spt} strategy unless stated otherwise. The scene contains $\Qd=2$ speech sources (VCTK~\cite{veaux2017vctk}) and $\Qn=2$ babble noise sources, at least $1$~m from any node, at $0$~dB mean \gls*{snr} (latent speech vs.\ noise), with sensor self-noise $20$~dB below the source power. The \gls*{cgls} observability follows \secref{ssec:oracle_validation}: one noise source is observed by all nodes, one by a single node.}

\change{Microphone signals are obtained by time-domain convolution of the latent source signals with the \glspl*{rir} and processed in a \gls*{wola} filter bank ($1024$-sample \gls*{dft}, $50$\% overlap, Hann window). Speech sources follow a $3$~s on-off cycle with randomized start times. The \glspl*{scm} are estimated by \gls*{vad}-based recursive averaging~\eqref{eq:expavg} with forgetting factor $\beta$, from a random initialization; an imperfect \gls*{vad} flips each per-source frame decision with probability $p_e$. Each $40$~s scenario is averaged over $20$ runs.}

\change{Under estimated \glspl*{scm}, a rank-$\cQ$ \gls*{gevd}-based \gls*{mwf}~\cite{serizel_low-rank_2014} replaces the plain \gls*{mwf} of \secref{sec:problem_statement} (standard for robustness against estimation errors), applied equally to the centralized, local, and \gls*{tidmwf} estimators. Fusion matrices and filters are recomputed every $N_{\mathrm{us}}=5$ frames ($160$~ms, cf.~\secref{sec:latency_complexity}) while the \glspl*{scm} update every frame. Performance is assessed by means of the \gls*{stoi}~\cite{taalAlgorithmIntelligibilityPrediction2011} over sliding windows, averaged over all $K$ nodes.}

\change{The third experiment tracks the \gls*{tidmwf} \gls*{stoi} over time, testing whether \gls*{scm} estimation errors accumulate through its cascaded estimation-fusion structure (upstream fusion-matrix errors propagating towards the root). It is compared to the centralized \gls*{gevdmwf} (estimating a single $M\times M$ \gls*{scm}), the local \gls*{gevdmwf}, and the unprocessed reference, at $\beta=0.99$ and $p_e\in\{0,0.05,0.1\}$.}

\begin{figure}[htbp]
  \centering
  \includegraphics[width=.9\columnwidth,trim={0 0 0 0},clip=false]{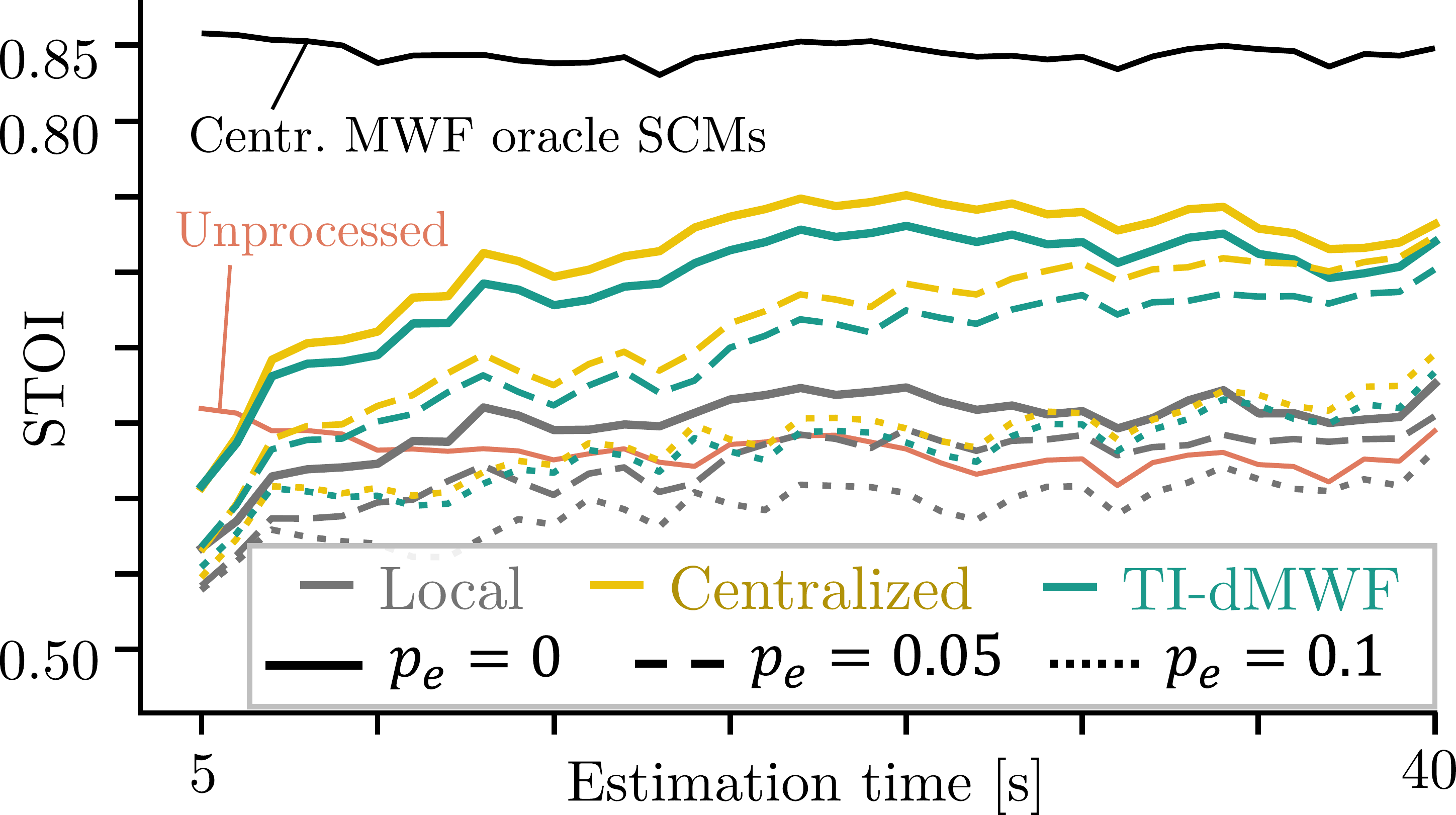}
  \caption{All-node mean \gls*{stoi} vs. time at $\beta=0.99$, over $20$ \gls*{cgls} scenarios, for \gls*{vad} error probabilities $p_e\in\{0,0.05,0.1\}$ (solid, dashed, dotted, respectively).}
  \label{fig:tidmwf:estscm}
\end{figure}

\change{\figref{fig:tidmwf:estscm} shows that the online estimated-\gls*{scm} filters reach a steady regime after an initial transient of about $15$~s. At $\beta=0.99$ and $p_e=0$, the \gls*{tidmwf} closely tracks the centralized \gls*{gevdmwf}, with a converged all-node \gls*{stoi} of $0.76$ versus $0.77$, both clearly exceeding the local \gls*{gevdmwf} ($0.66$) and the unprocessed reference ($0.63$). The \gls*{tidmwf} thus matches the centralized estimated-\gls*{scm} \gls*{mwf} to within a small margin, even though each of its nodes estimates only a small $\hMk[q]\times\hMk[q]$ \gls*{scm} rather than the single $M\times M$ centralized \gls*{scm}. \gls*{scm} estimation errors thus do not accumulate through the cascaded estimation-fusion structure, despite upstream fusion-matrix errors propagating towards the root. Robustness to an imperfect \gls*{vad} is shown by the $p_e=0.05$ (dashed) and $p_e=0.1$ (dotted) curves, for which the converged \gls*{tidmwf} \gls*{stoi} degrades monotonically to $0.74$ and $0.66$, respectively, while preserving the convergence behavior and the relative ordering of the estimators.}

\change{The fourth experiment assesses how the tree-pruning strategy affects \gls*{scm} estimation-error propagation. The same scenarios are processed with five strategies: \gls*{spt}, \gls*{mmut}, and \gls*{mst} on the $C=0.5$ ad-hoc topology, plus synthetic star and line topologies bounding the tree depth ($1$ for star, $K-1$ for line). The per-frame downstream cost is identical across strategies (any spanning tree of $K$ nodes has $K-1$ edges, each carrying a $\cQ$-dimensional fused signal), so the choice is purely a latency-vs-performance trade-off.}


\change{The converged all-node \gls*{tidmwf} \gls*{stoi} and mean tree depth per strategy ($20$ \gls*{cgls} scenarios, $\beta=0.99$, $p_e=0$) are summarized below; the tree-independent references are local $0.661$, centralized $0.785$, and unprocessed $0.626$.}

\begin{center}
\colorbox{black!8}{%
  \footnotesize\setlength{\tabcolsep}{4.5pt}\renewcommand{\arraystretch}{0.95}%
  \begin{tabular}{lccccc}
     & star & \gls*{spt} & \gls*{mmut} & \gls*{mst} & line \\
    \hline
    Mean depth & $1.0$ & $2.6$ & $3.1$ & $3.6$ & $4.0$ \\
    \gls*{stoi} & $0.792$ & $0.767$ & $0.767$ & $0.755$ & $0.757$ \\
  \end{tabular}%
}%
\end{center}

\change{The \gls*{tidmwf} performance varies only marginally, from $0.79$ \gls*{stoi} for the shallow star topology (depth $1$) to $0.76$ for the deeper \gls*{mst} and line topologies, a spread within the inter-scenario variability, with a mild trend favoring shallower trees (fewer cascaded fusion stages). Since the per-frame communication cost is identical across strategies, the choice is governed by latency, for which the \gls*{spt} is preferable as the minimum-depth tree for an arbitrary topology.}

\section{Conclusion}\label{sec:conclusion}

We have introduced the \gls*{tidmwf}, an algorithm for distributed signal estimation in topology-unconstrained \glspl*{wasn} that lets each node compute its node-specific centralized \gls*{mwf} by exchanging only low-dimensional fused signals, without iterative filter estimation, unlike state-of-the-art algorithms.
Simulations confirm that the \gls*{tidmwf} is optimal in \gls*{gls} scenarios, where each source is observed by all nodes or only one. Latency and complexity analyses show that the \gls*{tidmwf} retains centralized optimality in the \gls*{gls} setting with predictable latency, computational load, and communication cost, and degrades controllably when the \gls*{gls} assumption only approximately holds. Future work may relax the \gls*{gls} assumption and validate the algorithm in real-world scenarios \change{ beyond the present simulation study}, \change{and generalize the upstream signal construction~\eqref{eq:tidmwf:yuk} beyond the selection-matrix formulation to handle $\cQ>M_k$ at the root node}.

\section*{Appendix A: TI-dMWF Optimality}\label{sec:tidmwf_opt}

We first show that a leaf node $l\in\Kbq[k]$ estimating $\Euk^\T\cyk$ from its local signals $\yk[l]$ obtains an \gls*{lmmse} filter usable as a fusion matrix that compresses $\yk[l]$ into a lower-dimensional fused signal, which its downstream neighbor can then use to estimate $\Euk^\T\cyk$ without performance loss. Applying this recursively from the leaves to the root node $k$ yields the centralized \gls*{lmmse} estimator of $\Euk^\T\cyk$ at node $k$, achievable in practice by having each node transmit estimates of $\Euk^\T\yk$. The same process finally yields the centralized \gls*{lmmse} estimator for $\dk$ at node $k$.

For convenience, we rewrite the signal model as $\yk = \cyk + \uyk = \oHk\cxlat + \uyk$, where $\oHk\in\C[M_k][{\hQk}]$ is the steering matrix relative to all \textit{global} sources in the environment and $\cxlat\in\C[{\hQk}]$ is the vector of their latent source signals.
Consider the problem of estimating $\Euk^\T\cyk\in\C[{\hQk}]$ at a node $l\in\Kbq[k]$ based solely on its local sensor signals $\yk[l]$. The corresponding \gls*{lmmse} formulation is:

\vspace{-.5em}
\begin{align}
    \tW[l] &= \argmin[{\mathbf{W}_l\in\C[M_l][{\hQk}]}]
    \El[{
        \Euk^\T\cyk - \mathbf{W}_l^\Her\yk[l]
    }]\label{eq:proof_lmmse_leaf}\\
    &= \Rykyk[l]^{-1}\Rykcyq[l][k]\Euk,\label{eq:proof_mwf}
\end{align}

\noindent
where $\Rykcyq[l][k] = \E[{\yk[l]\cyk^\Her}] = \oHk[l]\cRxx\oHk^\Her$ with $\cRxx = \E[{\cxlat(\cxlat)^\Her}]$.
The \gls*{mwf}~\eqref{eq:proof_mwf} can be expanded using the matrix inversion lemma to obtain:

\vspace{-.5em}
\begin{align}
    \tW[l] &= \left(
        \Ruykuyq[l][l] + \oHk[l]\cRxx\oHk[l]^\Her
    \right)^{-1}
    \oHk[l]\cRxx\oHk^\Her\Euk\\
    &= \underbrace{\Ruykuyq[l][l]^{-1}\oHk[l]}_{M_l\times \hQk}
    \underbrace{
        \left(
            \mathbf{I}_{\hQk} - \Xk[l]^{-1}\oHk[l]^\Her\Ruykuyq[l][l]^{-1}\oHk[l]
        \right)\cRxx\oHk^\Her\Euk
    }_{=\Mk[lk]\in\C[{\hQk}][{\hQk}]},\label{eq:proof_solution_leaf}\\
    &\with \Xk[l] =
    \cRxx^{-1} + \oHk[l]^\Her\Ruykuyq[l][l]^{-1}\oHk[l],\label{eq:proof_solution_leaf_final}
\end{align}

\noindent
and where $\Ruykuyq[l][l] = \E[{\uyk[l]\uyk[l]^\Her}]$ is full-rank and square, thus invertible.
The term $\Ruykuyq[l][l]^{-1}\oHk[l]$ in~\eqref{eq:proof_solution_leaf} acts as a fusion matrix compressing $\yk[l]$ into a $\hQk$-dimensional signal, post-multiplied by $\Mk[lk]$ to obtain the \gls*{lmmse} filter $\tW[l]$.

Now consider an internal node $l'$ (direct downstream neighbor of $l$) estimating $\Euk^\T\cyk$ from $\yk[l']$ and the $\yk[l]$ transmitted by node $l$. The corresponding \gls*{lmmse} problem is:

\vspace{-.5em}
\begin{align}
    \tW[l'l] &= \argmin[{\mathbf{W}_{l'l}\in\C[(M_{l'}+M_{l})][{\hQk}]}]
    \El[{
        \Euk^\T\cyk - \mathbf{W}_{l'l}^\Her
        \begin{bmatrix}
            \yk[l']\\
            \yk[l]
        \end{bmatrix}
    }].
\end{align}

\noindent
Following a similar reasoning as~\eqref{eq:proof_lmmse_leaf}-\eqref{eq:proof_solution_leaf_final}, an alternative expression for $\tW[l'l]$ can be obtained:

\vspace{-.5em}
\begin{align}
    \tW[l'l] &=
        \begin{bmatrix}
            \Ruykuyq[l'][l']^{-1}\oHk[l']\\
            \Ruykuyq[l][l]^{-1}\oHk[l]
        \end{bmatrix}
    \mathbf{M}_{l'lk}.\label{eq:proof_solution_twonodes}\\
    \mathbf{M}_{l'lk} &= 
    \left(
        \I[{\hQk}] - \Xk[l'l]^{-1}\oHk[l'l]^\Her\Gam\oHk[l'l]
    \right)
    \cRxx\oHk^\Her\Euk\\
    \Xk[l'l]&=(\cRxx)^{-1} + \oHk[l'l]^\Her\Gam\oHk[l'l],
\end{align}

\noindent
with $\Ruykuyq[l'][l'] = \E[{\uyk[l']\uyk[l']^\Her}]$,
$\oHk[l'l] = [\oHk[l']^\T\:|\:\oHk[l]^\T]^\T$, and $\Gam=\bd[{\Ruykuyq[l'][l']^{-1},\Ruykuyq[l][l]^{-1}}]$.
As in~\eqref{eq:proof_solution_leaf}, the leftmost products $\Ruykuyq[l'][l']^{-1}\oHk[l']$ and $\Ruykuyq[l][l]^{-1}\oHk[l]$ in~\eqref{eq:proof_solution_twonodes} act as fusion matrices compressing $\yk[l']$ and $\yk[l]$ into $\hQk$-dimensional signals.

A fusion matrix is \textit{valid} (in the sense of this proof) if an \gls*{lmmse} problem using the fused signals yields the same result as one using the original signals. The matrices $\Ruykuyq[l'][l']^{-1}\oHk[l']$ and $\Ruykuyq[l][l]^{-1}\oHk[l]$ are valid, being the \gls*{lmmse} filters of nodes $l'$ and $l$ when estimating $\Euk^\T\cyk$ from $\yk[l']$ and $\yk[l]$ alone, cf.~\eqref{eq:proof_solution_leaf} (proof omitted).

Hence node $l'$'s estimate of $\Euk^\T\cyk$ from $\yk[l']$ and $\yk[l]$ is unchanged if node $l$ transmits the fused $(\Ruykuyq[l][l]^{-1}\oHk[l])^\Her\yk[l]\in\C[{\hQk}]$ instead of $\yk[l]\in\C[M_l]$. Moreover, post-multiplying $\Ruykuyq[l][l]^{-1}\oHk[l]$ by any invertible $\hQk\times \hQk$ matrix (e.g., $\Mk[lk]$ in~\eqref{eq:proof_solution_leaf} or $\mathbf{M}_{l'lk}$ in~\eqref{eq:proof_solution_twonodes}) gives an equally valid fusion matrix (proof omitted), so the estimate is likewise unchanged if $l$ transmits $\tW[l]^\Her\yk[l]\in\C[{\hQk}]$ instead of $\yk[l]$.

Formula~\eqref{eq:proof_solution_twonodes} extends to any number of cooperating nodes. In a tree, each \textit{leaf} node $l$ first computes a valid fusion matrix $\tW[l]$ from its local observation $\yk[l]$ as in~\eqref{eq:proof_solution_leaf}. Each \textit{internal} node $p$ then combines the fused signals from its upstream neighbors with $\yk[p]$ to compute its own \gls*{lmmse} filter for $\Euk^\T\cyk$, of the form~\eqref{eq:proof_solution_twonodes}, which serves as its fusion matrix to compress all available signals into a single $\hQk$-dimensional signal. Repeating this up to the root node $k$ yields the centralized \gls*{lmmse} estimator for $\Euk^\T\cyk$ there.

Since $\cyk$ is not directly available in practice, $\Euk^\T\cyk$ can be replaced by $\Euk^\T\yk$ (cf.~\eqref{eq:tidmwf:yuk}) in the \gls*{lmmse} estimation at all nodes but $k$ without changing the obtained filters since $\uyk$ is uncorrelated to $\uyk[l]$ by definition for any $l\in\Kbq[k]$.
Once node $k$ has received the same fused signals, the centralized \gls*{lmmse} estimator for $\dk=\Ek^\T\sk[]$ is likewise computed at $k$, with a filter analogous to~\eqref{eq:proof_solution_twonodes} (omitted for brevity). This concludes the proof of the optimality of the \gls*{tidmwf}.

\bibliographystyle{IEEEbib_mod}
\newpage
\bibliography{IEEEabrv,refs}

@STRING{IEEE_J_ASLP       = "{IEEE/ACM} Trans. Audio, Speech, Language Process."}

@STRING{IEEE_J_SP         = "{IEEE} Trans. Signal Process."}

@STRING{IEEE_J_SIPN       = "{IEEE} Trans. Signal Inf. Process. Netw."}

@STRING{IEEE_J_SENSOR     = "{IEEE} Sensors J."}

@STRING{IEEE_M_SP         = "{IEEE} Signal Process. Mag."}

@STRING{JASA       = "{Journal of the Acoustical Society of America}"}

@STRING{WCMC       = "{Wireless Communications and Mobile Computing}"}

@STRING{BSTJ       = "{Bell System Technical Journal}"}

@STRING{PROC_ICASSP       = "{Proceedings of the IEEE International Conference on Acoustics, Speech, and Signal Processing}"}

@STRING{PROC_EUSIPCO       = "{Proceedings of the European Signal Processing Conference}"}

@STRING{PROC_SCVT       = "{Proceedings of the IEEE Symposium on Communications and Vehicular Technology}"}

@STRING{PROC_AMS       = "{Proceedings of the American Mathematical Society}"}

@STRING{PROC_ONCON       = "{Proceedings of the Industrial Electronics Society Annual On-Line Conference}"}

@inproceedings{bertrand_applications_2011,
title = {{Applications and trends in wireless acoustic sensor networks: {A} signal processing perspective}},
booktitle = PROC_SCVT,
author = {Bertrand, A.},
year = {2011},
pages = {1--6},
}

@article{bertrand_distributed_2010,
title = {{Distributed Adaptive Node-Specific Signal Estimation in Fully Connected Sensor Networks—{Part} {I}: Sequential Node Updating}},
volume = {58},
number = {10},
journal = IEEE_J_SP,
author = {Bertrand, A. and Moonen, M.},
year = {2010},
pages = {5277--5291},
}

@article{szurley_topology-independent_2017,
title = {{Topology-Independent Distributed Adaptive Node-Specific Signal Estimation in Wireless Sensor Networks}},
volume = {3},
number = {1},
journal = IEEE_J_SIPN,
author = {Szurley, J. and Bertrand, A. and Moonen, M.},
year = {2017},
pages = {130--144},
}

@article{serizel_low-rank_2014,
title = {{Low-rank Approximation Based Multichannel {Wiener} Filter Algorithms for Noise Reduction with Application in Cochlear Implants}},
volume = {22},
number = {4},
journal = IEEE_J_ASLP,
author = {Serizel, R. and Moonen, M. and Van Dijk, B. and Wouters, J.},
year = {2014},
pages = {785--799}
}

@article{hassani_gevd-based_2016,
title = {{GEVD}-Based Low-Rank Approximation for Distributed Adaptive Node-Specific Signal Estimation in Wireless Sensor Networks},
volume = {64},
number = {10},
journal = IEEE_J_SP,
author = {Hassani, A. and Bertrand, A. and Moonen, M.},
year = {2016},
pages = {2557--2572},
}

@article{cobos_survey_2017,
title = {{A {Survey} of {Sound} {Source} {Localization} {Methods} in {Wireless} {Acoustic} {Sensor} {Networks}}},
volume = {2017},
journal = WCMC,
author = {Cobos, Maximo and Antonacci, Fabio and Alexandridis, Anastasios and Mouchtaris, Athanasios and Lee, Bowon},
year = {2017},
pages = {1--24},
}

@article{didier2025fastconvergingdistributedsignalestimation,
title = {{Fast-Converging Distributed Signal Estimation in Topology-Unconstrained Wireless Acoustic Sensor Networks}},
author = {Didier, Paul and van Waterschoot, Toon and Doclo, Simon and Bitzer, J{\"o}rg and Moonen, Marc},
year = {2025},
journal = {arXiv preprint arXiv:2506.02797},
note = {Submitted for publication in } # IEEE_J_SIPN,
archivePrefix = {arXiv},
eprint = {2506.02797},
primaryClass = {eess.AS},
}

@article{prim1957shortest,
title={{Shortest connection networks and some generalizations}},
author={Prim, Robert Clay},
journal=BSTJ,
volume={36},
number={6},
pages={1389--1401},
year={1957},
}

@article{dijkstra1959note,
title={{A note on two problems in connexion with graphs}},
author={Dijkstra, Edsger W.},
journal={{Numerische Mathematik}},
volume={1},
number={1},
pages={269--271},
year={1959},
}

@inproceedings{didier2024tigevddanse,
title={{Topology-Independent {GEVD}-Based Distributed Adaptive Node-Specific Signal Estimation in Ad-Hoc Wireless Acoustic Sensor Networks}},
booktitle = PROC_EUSIPCO,
author={Didier, Paul and van Waterschoot, Toon and Moonen, Marc},
year = {2024},
pages = {1--5},
}

@article{kruskal1956shortest,
title={{On the shortest spanning subtree of a graph and the traveling salesman problem}},
author={Kruskal, Joseph B},
journal=PROC_AMS,
volume={7},
number={1},
pages={48--50},
year={1956}
}

@inproceedings{musluoglu_distributed_2023,
title = {{A {Distributed} {Adaptive} {Algorithm} for {Node}-{Specific} {Signal} {Fusion} {Problems} in {Wireless} {Sensor} {Networks}}},
booktitle = PROC_EUSIPCO,
author = {Musluoglu, Cem Ates and Bertrand, Alexander},
year = {2023},
pages = {1654--1658},
doi = {10.23919/EUSIPCO58844.2023.10289843},
}

@article{doclo_reduced-bandwidth_2009,
title = {{Reduced-{Bandwidth} and {Distributed} {MWF}-{Based} {Noise} {Reduction} {Algorithms} for {Binaural} {Hearing} {Aids}}},
volume = {17},
number = {1},
journal = IEEE_J_ASLP,
author = {Doclo, S. and Moonen, M. and Van den Bogaert, T. and Wouters, J.},
year = {2009},
pages = {38--51},
}

@inproceedings{scheiblerPyroomacousticsPythonPackage2018,
title = {{Pyroomacoustics: {{A Python Package}} for {{Audio Room Simulation}} and {{Array Processing Algorithms}}}},
booktitle = PROC_ICASSP,
author = {Scheibler, Robin and Bezzam, Eric and Dokmani{\'c}, Ivan},
year = {2018},
pages = {351--355},
}

@techreport{veaux2017vctk,
author       = {Veaux, C. and Yamagishi, J. and MacDonald, K.},
title        = {{CSTR VCTK Corpus: English Multi-speaker Corpus for CSTR Voice Cloning Toolkit}},
institution  = {The Centre for Speech Technology Research (CSTR), University of Edinburgh},
year         = {2017},
note         = {Accessed: August 14, 2025}
}

@article{taalAlgorithmIntelligibilityPrediction2011,
title = {{An {{Algorithm}} for {{Intelligibility Prediction}} of {{Time}}--{{Frequency Weighted Noisy Speech}}}},
author = {Taal, Cees H. and Hendriks, Richard C. and Heusdens, Richard and Jensen, Jesper},
year = {2011},
journal = IEEE_J_ASLP,
volume = {19},
number = {7},
pages = {2125--2136},
}

@misc{didier2025dmwf,
title = {{Distributed Multichannel Wiener Filtering for Wireless Acoustic Sensor Networks}},
author = {Didier, Paul and van Waterschoot, Toon and Doclo, Simon and Bitzer, J{\"o}rg and Behmandpoor, Pourya and Gode, Henri and Moonen, Marc},
year = {2025},
note = {Submitted for publication in } # IEEE_J_ASLP,
}

@inproceedings{natarajanComprehensiveReviewBeamformingBased2023,
title = {{A {{Comprehensive Review}} of {{Beamforming-Based Speech Enhancement Techniques}}, {{IoT}}, and {{Smart City Applications}}}},
booktitle = PROC_ONCON,
author = {Natarajan, Sureshkumar and {Al-Haddad}, Syed Abdul Rahman and Ahmad, Faisul Arif and Hassan, Mohd Khair and Kamil, Raja and Azrad, Syaril and Macleans, June Francis},
year = {2023},
pages = {1--6}
}

@article{griffinLocalizingMultipleAudio2015,
title = {{Localizing Multiple Audio Sources in a Wireless Acoustic Sensor Network}},
author = {Griffin, Anthony and Alexandridis, Anastasios and Pavlidi, Despoina and Mastorakis, Yiannis and Mouchtaris, Athanasios},
year = {2015},
journal = {Signal Processing},
series = {Special {{Issue}} on Ad Hoc Microphone Arrays and Wireless Acoustic Sensor Networks},
volume = {107},
pages = {54--67}
}

@article{changDistributedKalmanFiltering2023,
title = {{Distributed {{Kalman Filtering}} for {{Speech Dereverberation}} and {{Noise Reduction}} in {{Acoustic Sensor Networks}}}},
author = {Chang, Ruijiang and Chen, Zhe and Yin, Fuliang},
year = {2023},
journal = IEEE_J_SENSOR,
volume = {23},
number = {24},
pages = {31027--31037}
}

@inproceedings{lohmannDereverberationAcousticSensor2023,
title = {{Dereverberation in {{Acoustic Sensor Networks Using}} Weighted {{Prediction Error}} with {{Microphone-Dependent Prediction Delays}}}},
booktitle = PROC_ICASSP,
author = {Lohmann, Anselm and {van Waterschoot}, Toon and Bitzer, Joerg and Doclo, Simon},
year = {2023},
pages = {1--5}
}

@article{musluogluUnifiedAlgorithmicFramework2023,
title = {{A {{Unified Algorithmic Framework}} for {{Distributed Adaptive Signal}} and {{Feature Fusion Problems}} -- {{Part I}}: {{Algorithm Derivation}}}},
author = {Musluoglu, Cem Ates and Bertrand, Alexander},
year = {2023},
journal = IEEE_J_SP,
volume = {71},
eprint = {2208.08867},
primaryclass = {eess},
pages = {1863--1878},
archiveprefix = {arXiv},
doi = {10.1109/TSP.2023.3275272},
}

@article{soudenGaussianModelBasedMultichannel2010a,
title = {{Gaussian {{Model-Based Multichannel Speech Presence Probability}}}},
author = {Souden, Mehrez and {Jingdong Chen} and Benesty, Jacob and Affes, Sofi{\`e}ne},
year = {2010},
journal = IEEE_J_ASLP,
volume = {18},
number = {5},
pages = {1072--1077},
}

@inproceedings{tammenDNNBasedSpeechPresence2020,
title = {{DNN-Based Speech Presence Probability Estimation} for {Multi-Frame Single-Microphone Speech Enhancement}},
booktitle = PROC_ICASSP,
author = {Tammen, Marvin and Fischer, Dorte and Meyer, Bernd T. and Doclo, Simon},
year = {2020},
pages = {191--195},
}

@article{zhaoModelbasedDistributedNode2020,
title = {{Model-Based Distributed Node Clustering and Multi-Speaker Speech Presence Probability Estimation in Wireless Acoustic Sensor Networks}},
author = {Zhao, Yingke and Nielsen, Jesper Kj{\ae}r and Chen, Jingdong and Christensen, Mads Gr{\ae}sb{\o}ll},
year = {2020},
journal = JASA,
volume = {147},
number = {6},
pages = {4189--4201},
}

@misc{bahariDistributedMultiSpeakerVoice2017a,
title = {{Distributed {{Multi-Speaker Voice Activity Detection}} for {{Wireless Acoustic Sensor Networks}}}},
author = {Bahari, Mohamad Hasan and Hamaidi, L. Khadidja and Muma, Michael and {Plata-Chaves}, Jorge and Moonen, Marc and Zoubir, Abdelhak M. and Bertrand, Alexander},
year = {2017},
number = {arXiv:1703.05782},
eprint = {1703.05782},
primaryclass = {stat},
archiveprefix = {arXiv},
}

@article{brcichDetectionSourcesUsing2002,
title = {{Detection of Sources Using Bootstrap Techniques}},
author = {Brcich, R.F. and Zoubir, A.M. and Pelin, P.},
year = {2002},
journal = IEEE_J_SP,
volume = {50},
number = {2},
pages = {206--215},
}

@article{stoicaModelorderSelectionReview2004,
title = {{Model-Order Selection: A Review of Information Criterion Rules}},
author = {Stoica, P. and Selen, Y.},
year = {2004},
journal = IEEE_M_SP,
volume = {21},
number = {4},
pages = {36--47},
}

@article{waxDetectionNumberSignals2021,
title = {{Detection of the {{Number}} of {{Signals}} by {{Signal Subspace Matching}}}},
author = {Wax, Mati and Adler, Amir},
year = {2021},
journal = IEEE_J_SP,
volume = {69},
pages = {973--985},
}

@article{williamsUsingSphericityTest1990,
title = {{Using the Sphericity Test for Source Detection with Narrow-Band Passive Arrays}},
author = {Williams, D.B. and Johnson, D.H.},
year = {1990},
journal = IEEE_J_ASLP,
volume = {38},
number = {11},
pages = {2008--2014},
}

@book{benestySpeechEnhancement2006,
  title = {Speech {{Enhancement}}},
  author = {Benesty, Jacob and Makino, Shoji and Chen, Jingdong},
  year = {2006},
  month = mar,
  publisher = {Springer Science \& Business Media},
}

\vfill\pagebreak

\end{document}